%% file: main.tex
\documentclass[sigconf]{acmart}

\AtBeginDocument{%
  \providecommand\BibTeX{{%
    \normalfont B\kern-0.5em{\scshape i\kern-0.25em b}\kern-0.8em\TeX}}}

\usepackage{booktabs}
\usepackage{wrapfig}
\usepackage{listings}
\usepackage{caption}
\usepackage{hyperref}
\usepackage{multirow}
\usepackage{balance}
\usepackage{enumitem}
\usepackage{framed}
\usepackage{xcolor}
\usepackage{tabularx}
\usepackage{subcaption}

\copyrightyear{2025}
\acmYear{2025}
\setcopyright{acmlicensed}\acmConference[DIS '25]{Designing Interactive Systems Conference}{July 5--9, 2025}{Funchal, Portugal}
\acmBooktitle{Designing Interactive Systems Conference (DIS '25), July 5--9, 2025, Funchal, Portugal}
\acmDOI{10.1145/3715336.3735839}
\acmISBN{979-8-4007-1485-6/2025/07}

\newcolumntype{L}[1]{>{\raggedright\let\newline\\\arraybackslash\hspace{0pt}}m{#1}}
\newcolumntype{C}[1]{>{\centering\let\newline\\\arraybackslash\hspace{0pt}}m{#1}}
\newcolumntype{R}[1]{>{\raggedleft\let\newline\\\arraybackslash\hspace{0pt}}m{#1}}

\newcommand{\partis}{{participants}}

\newcommand{\Partis}{{Participants}}

\newcommand{\nblind}{{$12$}}
\newcommand{\nsight}{{$12$}}

\newcommand{\sysname}{{ToPSen}}

\newcommand{\fx}[1]{{\textcolor{black}{#1}}}

\begin{document}


 \title[ToPSen: Task-Oriented Priming and Sensory Alignment for Comparing Coding Strategies]
 {ToPSen: Task-Oriented Priming and Sensory Alignment for Comparing Coding Strategies Between Sighted and Blind Programmers}
 
 %
 %
 %
 
\author{Md Ehtesham-Ul-Haque}
\affiliation{%
  \institution{Pennsylvania State University}
  \city{University Park}
  \state{Pennsylvania}
  \country{USA}
}
\email{mfe5232@psu.edu}

\author{Syed Masum Billah}
\affiliation{%
  \institution{Pennsylvania State University}
  \city{University Park}
  \state{Pennsylvania}
  \country{USA}
}
\email{sbillah@psu.edu}



\input{0_abstract}



\begin{teaserfigure}
    \centering
  \includegraphics[width=0.99\textwidth]{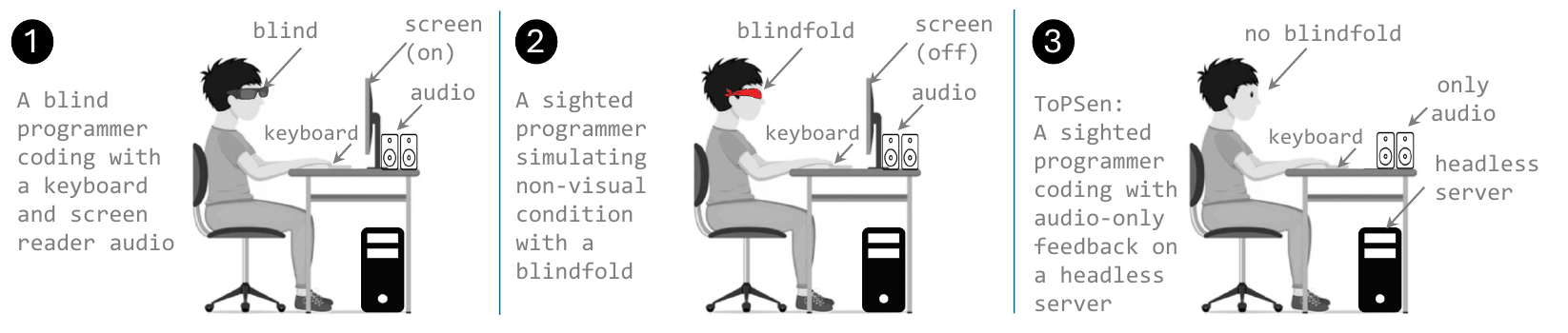}
  \caption{
    An illustration comparing non-visual coding practices between blind and sighted programmers. 
    \textbf{Left (1)}: A traditional setup for blind programmers where they use a keyboard for input and audio readout (from a screen reader) for output.
    \textbf{Center (2)}: A traditional simulation approach where sighted programmers wear a blindfold and use keyboard input with audio feedback, while the screen is turned off to simulate blindness.
    \textbf{Right (3)}: The \textit{\textbf{T}ask-\textbf{O}riented \textbf{P}riming and \textbf{Sen}sory Alignment} \textit{(ToPSen)} setup, where sighted programmers do not wear a blindfold in order to simulate blindness, instead, they are primed to code under real-world conditions, such as writing Python scripts on a headless server that does not have a graphical interface but only audio readout. 
%
  }
\Description{A comparative illustration showing three different programming setups, labeled 1, 2, and 3, each depicted with a stick figure programmer sitting at a desk with computer equipment.
Setup 1 shows a blind programmer using standard assistive technology: they wear glasses, use a keyboard for input, and have speakers for screen reader audio output. The monitor is turned on.
Setup 2 demonstrates a traditional blindness simulation approach: a sighted programmer wearing a red blindfold, using a keyboard, with speakers for audio output. The monitor is explicitly marked as turned off.
Setup 3 presents the ToPSen approach: a sighted programmer without any blindfold, using a keyboard and speakers for audio output, working with a headless server setup. The programmer appears identical to the others, but without glasses or a blindfold.
Each setup is drawn in a simple, grayscale style with clear labels for equipment (keyboard, screen, audio) and programmer status. The three setups are separated by vertical lines and are of equal size. The figures are shown in profile view, all seated in office chairs at similar desks with desktop computer towers beneath.
Above each setup is descriptive text explaining the specific configuration and purpose: the first shows a blind programmer's typical setup, the second shows a blindfolded simulation approach, and the third introduces the ToPSen method using audio-only feedback on a headless server.}
  \label{fig:teaser}
\end{teaserfigure}


\maketitle

\input{1_intro}

\input{2_related_work}
\input{3_design}
\input{4_findings}
\input{5_discussion}
\input{6_conclusion}

\begin{acks}
We thank anonymous reviewers for their insightful feedback. 
This work was partially supported by Seed Grant No.\ 150000008684 from the College of Information Sciences and Technology at Pennsylvania State University.
\end{acks}

\balance
\bibliographystyle{ACM-Reference-Format}
\bibliography{local, Bibliography, Bibliography2, Bibliography3}

\end{document}

%% file: 0_abstract.tex
\begin{abstract}
This paper examines how the coding strategies of sighted and blind programmers differ when working with audio feedback alone. The goal is to identify challenges in mixed-ability collaboration, particularly when sighted programmers work with blind peers or teach programming to blind students. To overcome limitations of traditional blindness simulation studies, we proposed Task-Oriented Priming and Sensory Alignment (ToPSen), a design framework that reframes sensory constraints as technical requirements rather than as a disability. Through a study of 12 blind and 12 sighted participants coding non-visually, we found that expert blind programmers maintain more accurate mental models and process more information in working memory than sighted programmers using ToPSen. Our analysis revealed that blind and sighted programmers process structural information differently, exposing gaps in current IDE designs. These insights inform our guidelines for improving the accessibility of programming tools and fostering effective mixed-ability collaboration.
\end{abstract}

%% file: 1_intro.tex
\section{Introduction}
Programmers face significant cognitive demands when writing code, which stem from both intrinsic and extraneous sources~\cite{sweller2011cognitive, sweller1988cognitive}. 
They need to perform complex mental operations such as abstract thinking, problem decomposition, and code comprehension, often requiring them to process multiple interrelated elements simultaneously to grasp their connections and behaviors~\cite{bederson1999does}. These constitute the intrinsic cognitive load.
Beyond intrinsic load, programmers must also handle extraneous cognitive demands --  tasks necessary for coding but not directly related to problem-solving. These include perceiving and manipulating code text through sensory and motor channels, memorizing syntax rules, and identifying and fixing syntax errors~\cite{sweller2011cognitive, stefik2013empirical, kaijanaho2015evidence}.

When using identical tools and environments, programmers' performance varies primarily with their programming skills~\cite{leppink2013development}. 
However, different perceptual channels can fundamentally alter how programmers interact with code, making it particularly challenging to compare the performance of sighted and blind programmers.
This is because sighted programmers perceive code visually, while blind programmers interact with text through auditory or tactile channels, typically using text-to-speech technologies, as shown in Fig.~\ref{fig:teaser}.1. 
For identical coding tasks, this fundamental difference in sensory perception creates varying extraneous cognitive loads that affect blind programmers' coding performance and efficiency~\cite{kane2014tracking, stefik2011design, Potluri2018CodeTalk, Albusays2017Interviews, mountapmbeme2022addressing, ehtesham2022gridcoding}. 
Yet researchers have not quantified how these perceptual differences influence coding performance and strategies.

This gap in understanding creates several challenges. To our knowledge, no studies directly compare sighted and blind programmers under controlled sensory and task conditions, leaving their performance differences unquantified. 
Consequently, we lack metrics for how much coding tools need modification to accommodate blind programmers effectively. 
Conversely, we do not fully understand what adaptations sighted programmers need to collaborate with blind colleagues or teach blind students. 
These unknowns impede the development of truly inclusive coding environments.

Our research addresses these challenges by controlling perceptual channels -- a key source of extraneous cognitive load -- for both sighted and blind programmers while standardizing coding tools, environments, and tasks. 

Traditional research controls perceptual channels through disability simulation. For non-visual programming, researchers typically blindfold sighted programmers to match blind programmers' experience (Fig.~\ref{fig:teaser}.2). However, these simulations face criticism due to their unproven effectiveness and negative impacts. Critics note that disability simulations often induce feelings of helplessness and overemphasize the initial shock of vision loss rather than capturing authentic experiences~\cite{bennett2019promise, flower2007meta, silverman2015perils}.

\subsection{\textit{\textbf{T}ask-\textbf{O}riented \textbf{P}riming and \textbf{Sen}sory Alignment} \textit{(ToPSen)} Design}
\label{sec:topsen}

To address these limitations, we propose \textit{\textbf{T}ask-\textbf{O}riented \textbf{P}riming and \textbf{Sen}sory Alignment} \textit{(ToPSen)} design (Fig.~\ref{fig:teaser}.3).
\sysname{} builds on three key principles. First, task-oriented priming that presents non-visual coding as a technical requirement -- similar to programming on a headless server with only audio feedback. Second, sensory alignment that ensures both sighted and blind programmers use identical auditory feedback through text-to-speech output. 
Third, carefully scoped programming tasks under 30 lines that allow both groups to participate without extensive training in assistive technologies.

Our approach fundamentally differs from traditional blindness simulation. 
Rather than blindfolding sighted programmers, ToPSen allows them to use vision for non-task activities -- like looking through windows or around their workspace -- while concentrating on audio programming.
This shift encourages sighted programmers to develop mental models for non-visual coding and approach programming as a constrained problem-solving task. By aligning both groups' sensory channels to audio, we enable direct comparison of coding strategies while avoiding the emotional distress typically associated with disability simulation.

\subsection{Research Questions}
\label{sec:rq}
ToPSen enables us to examine how different sensory modalities affect programming efficiency and strategy. We investigate:

\begin{itemize}
    \item[\textbf{\textit{RQ1}}] \textit{How do coding strategies compare between sighted programmers constrained to auditory perception and blind programmers reliant on auditory input during common coding tasks?}   
    \item [\textbf{\textit{RQ2}}] \textit{How do perceptual experiences shape programmers' mental models during auditory programming, and how do these models compare between sighted and blind programmers?}    

\end{itemize}

RQ1 reveals how both groups adapt their programming strategies to auditory input and manage extraneous cognitive load. These insights will guide IDE developers in designing tools that reduce cognitive burden during auditory programming.
RQ2 examines how different perceptual experiences shape programmers' mental models of code in controlled environments. This question also explores whether sighted programmers' visual perception creates unconscious biases that affect their ability to teach or collaborate with blind colleagues.
Such biases exist in other contexts, such as describing visual scenes~\cite{xie2023two}.

\subsection{Key Findings}
To answer these questions, we conducted a study with 12 blind and 12 sighted programmers completing identical coding tasks using a plain-text editor with audio feedback.
We found that both groups developed mental models by listening to code line-by-line.
Expert blind programmers successfully manage both intrinsic and extraneous cognitive loads by actively tracking code structure, staying aware of cursor location, and performing code reviews frequently to keep a consistent mental model. 
In contrast, sighted programmers using ToPSen (whom we call ToPSen programmers) focus primarily on syntax and semantics without paying much attention to code structure, level, and editing location. 
Novice programmers in both groups struggle with high intrinsic cognitive load, regardless of their familiarity with audio processing.

\subsection{Key Design Implications}
Based on these findings, we identify essential components of the mental models for auditory programming. 
Our analysis reveals that sighted programmers often overlook structural and positional information -- elements they typically process visually without conscious effort. 
This gap in attention creates communication barriers between blind and sighted programmers. We propose design guidelines for inclusive programming environments that bridge these differences and enhance mixed-ability collaboration.

\subsection{Key Contributions}
Our research makes four key contributions:

\begin{itemize}

\item We introduce ToPSen (Task-Oriented Priming and Sensory Alignment), a novel approach that reframes auditory programming as a technical challenge for sighted programmers rather than a disability simulation, enabling direct comparison of coding strategies between blind and sighted programmers.
\item We present empirical findings from 24 programmers (12 blind, 12 sighted) using ToPSen design, documenting how both groups approach coding tasks with audio-only feedback
\item We identify core components of auditory programming mental models and reveal how blind and sighted programmers differ in applying these components.
\item We outline design opportunities to support blind programmers' mental models and facilitate effective collaboration between programmers with different perceptual abilities.
\end{itemize}

%% file: 2_related_work.tex
\section{Background and Related Work}
In this section, we review prior work on programming challenges for blind programmers, examine how coding relies on different sensory inputs, and discuss critiques of traditional disability simulation methods. We then position our work within this context.

\subsection{Sighted vs. Blind Programming Experience}
Sighted programmers rely on feature-rich programming environments that provide visual cues~\cite{blinddeveloper, Albusays2017Interviews}. Highlighted keywords and matching brackets help them understand code structure and reduce extraneous cognitive load~\cite{sweller2011cognitive}. Programming languages often incorporate visual elements (e.g., indentation levels in Python) that enable sighted programmers to quickly parse code structure~\cite{blinddeveloper}. Moreover, sighted programmers can process multiple programming statements simultaneously through parallel visual perception.

In contrast, blind programmers use auditory channels to process code sequentially, one statement at a time~\cite{Albusays2017Interviews}. The visual cues in feature-rich environments remain inaccessible to them~\cite{Potluri2018CodeTalk}, leading them to prefer plain text editors~\cite{blinddeveloper, Albusays2016Eliciting, Albusays2017Interviews}. Blind programmers often must listen to code character by character to understand structure (e.g., hearing indentation as \texttt{\small space, space, space..}), which increases cognitive load~\cite{sweller2011cognitive, blinddeveloper}. 

Researchers have explored alternative code representations to support blind programmers, including tree structures~\cite{smith2000java, baker2015struct, Schanzer2019Accessible}, list structures~\cite{Potluri2018CodeTalk}, grid structures~\cite{ehtesham2022gridcoding}, and auditory cues~\cite{Potluri2018CodeTalk, stefik2009sodbeans, stefik2013empirical}. While extensive research compares sighted and blind users in non-programming tasks like wayfinding, spatial-cognitive mapping, and audio processing~\cite{passini1990spatio, listening_rate_pvi}, few studies directly compare their programming approaches. Armaly et al.~\cite{armaly2017comparison} found that blind programmers use strategies similar to sighted programmers when reading and understanding code snippets. 

However, research lacks comprehensive comparisons of performance and strategies between sighted and blind programmers across varied coding tasks. These tasks include code reading, navigation, comprehension, writing, and error correction. We particularly need studies in controlled conditions where both groups use identical sensory channels while maintaining access to their primary perceptual modes. This gap limits our understanding of how sensory adaptations affect programming activities --- a gap our research aims to address.

\subsection{Auditory Processing Abilities in Varying Degree of Blindness}
Research shows that blind individuals process auditory information at significantly higher rates than sighted individuals, demonstrating the brain's plasticity~\cite{listening_rate_pvi, plasticity_2009}. This enhanced listening ability reflects fundamental differences in how blind and sighted individuals' brains process auditory information~\cite{plasticity_2004}. 

Brain scan studies reveal that blind individuals engage their visual cortex -- a major brain region traditionally thought to process only visual stimuli -- for various cognitive processes~\cite{visual_cortex_2002, visual_cortex_2000, plasticity_2004, kolarik2020accuracy}. The processing capacity varies with blindness onset and duration. Congenitally blind individuals develop sophisticated spatial skills through non-visual cues, with their visual cortex actively supporting language processing~\cite{bedny2011language}. Those with early blindness show enhanced auditory and memory performance, particularly in verbal tasks~\cite{gougoux2005functional, roder2001auditory}. Late-blind individuals, despite initial adaptation challenges, learn to repurpose their visual brain areas for tactile and auditory processing~\cite{burton2003visual, pasqualotto2012role, voss2008differential}. 

Even temporary visual deprivation in sighted individuals can trigger rapid neuroadaptive changes, highlighting the brain's dynamic sensory processing capabilities~\cite{kauffman2002braille, merabet2008rapid}. Our research examines how varying levels of auditory adaptation affect coding tasks and how the absence of visual cues shapes programming performance and strategies.

\subsection{Disability Simulation, Empathy Building, and Alternatives}
Disability simulation -- an approach that attempts to foster empathy by having non-disabled people temporarily experience disability -- faces substantial criticism from researchers and advocates. Studies reveal both a lack of effectiveness data and potential negative impacts on participants, who often focus on feelings of helplessness rather than understanding the lived experience of disability~\cite{flower2007meta, silverman2015perils}. Research suggests that skill mastery and meaningful interaction with disabled individuals offer more accurate and respectful paths to understanding~\cite{silverman2015perils}.

Simulation-based design approaches introduce additional problems~\cite{bennett2019promise}. Designers may inadvertently reinforce power imbalances between themselves and those without design backgrounds and reduce disabled experiences to spectacles. These issues can compromise authentic disability experiences while privileging non-disabled designers' perspectives without sufficient accountability to the diverse communities they claim to represent.

Advocates propose moving from attempts to ``be like'' disabled individuals to ``being with'' them, emphasizing meaningful interactions over simulations. 
Our design approach, \sysname{}, detailed in Sec.~\ref{sec:topsen}, diverges from traditional disability simulations by focusing on objective task performance and technical challenges rather than subjective empathy building. We draw inspiration from situational impairments -- temporary difficulties in computer use caused by environmental factors (like lens glare or carrying shopping bags with both hands), medical conditions (such as post-eye surgery recovery), or injuries (like fractured bones) that can affect anyone. 
Our design challenges us to create authentic situations where participants remain consistently impaired throughout their tasks. We found this authenticity in a common programming scenario: writing scripts on a remote, headless server that provides only text-to-speech feedback over the internet. This scenario forms the foundation of our study described next.

\subsection{Mental Model of Programming}
\fx{
Mental models are internal cognitive representations that help programmers understand and predict system behavior. They play a central role in code comprehension, debugging, and navigation, enabling programmers to simulate execution and reason about structure~\cite{norman1983some, gentner1983mental}. Expert programmers typically build rich models that include syntax, control flow, and structural cues, while novices often struggle to maintain such coherence~\cite{robins2003learning, heinonen2023synthesizing}. In auditory or non-visual environments, mental models become even more critical due to the lack of spatial cues and the transient nature of audio~\cite{bragg2018listening}. Prior work shows that structured auditory tools, such as tree-based representations, can support users by making implicit code structure explicit and reducing cognitive load~\cite{Potluri2018CodeTalk}.
}

%% file: 3_design.tex
\section{Study: Comparison of Coding Strategies Between Sighted and Blind Programmers}

\begin{table*}[!t]
\small
\begin{tabular}{l p{1.0cm}  C{2.5cm} C{1.5cm}   C{4cm} C{3cm}}
\multirow{2}{*}{\textbf{ID}}	& \textbf{Age range}/	& \multirow{2}{*}{\textbf{Vision Status}} &  \textbf{Coding} &  \multirow{2}{*}{\textbf{IDE Used}} & \multirow{2}{*}{\textbf{Profession}} \\
& \textbf{Sex}   &                                     &          \textbf{Skill}  &                      &      \\ 
\toprule
B1	&   36-40/F	& Blind since 20        	  &     Expert & Notepad, Notepad++ & Distance Learner \\ \hline 
B2	&   71-75/M	& Congenital blindness         &     Expert & Notepad, Notepad++, EdSharp     & Music Teacher \\ \hline
B3	&   26-30/M	&  Congenital blindness  	  &     Expert & Notepad, Visual Studio Code & Distance Learner\\\hline 
B4	&   36-40/M	& Congenital blindness	          &		Beginner & Notepad, vi  & Business Entrepreneur\\ \hline
B5	&   46-50/M	& Congenital blindness	       	  &     Beginner & Notepad++ & Network Administrator \\ \hline
B6	&   76-80/M	& Congenital blindness	       	  &     Beginner & Notepad, EdSharp & Self-employed\\\hline 
B7  &   36-40/M	& Blind since 23        	  &     Expert & Notepad, Notepad++        & IT Instructor \\ \hline
B8	&   51-55/M	& Congenital blindness	       	  &     Expert & Notepad++, Visual Studio Code & Software Developer\\\hline 
B9	&   66-70/M	& Congenital blindness	       	  &     Beginner & Notepad, EdSharp & AT Instructor\\\hline 
B10	&   66-70/M	& Congenital	blindness       	  &     Expert & Notepad++, Visual Studio Code & Engineer\\\hline 
B11	&   56-60/M	& Blind since 16	       	  &     Expert & Notepad, EdSharp & Engineer\\\hline 
B12	&   41-45/M	& Congenital	blindness       	  &     Expert & Notepad++ & AT Instructor\\
\bottomrule
\bottomrule
S1	&	26-30/M	& Sighted  	  &     Beginner & Atom, Jupyter Notebook & Graduate Student\\ \hline
S2	&	26-30/M	& Sighted	  &     Expert & Visual Studio, Jupyter Notebook & Graduate Student \\ \hline
S3	&   21-25/M	& Sighted  	  &     Beginner & Visual Studio, Codeblocks & Graduate Student \\ \hline
S4	&	21-25/M	& Sighted  	  &     Beginner & Notepad, Sublime Text & Graduate Student\\ \hline
S5	&   26-30/M	& Sighted  	  &     Expert & Visual Studio Code, Jupyter Notebook & AI Researcher \\ \hline
S6	&   26-30/M	& Sighted     &     Beginner & Codeblocks, Visual Studio & Graduate Student\\ \hline
S7	&   26-30/M	& Sighted  	  &     Expert  & Visual Studio Code, Intellij & Software Engineer\\\hline 
S8	&   26-30/M	& Sighted  	  &     Expert  & Visual Studio Code, Jupyter Notebook & Teaching Assistant\\ \hline
S9	&   35-40/M	& Sighted  	  &     Expert  & Visual Studio Code, Intellij, Google Colab, Notepad, Notepad++, vi & Programming Instructor\\ \hline
S10	&   26-30/M	& Sighted  	  &     Expert  & Visual Studio Code, Google Colab & Teaching Assistant\\ \hline
S11	&   26-30/M	& Sighted  	  &     Expert  & Visual Studio Code, Google Colab, Notepad++ & Teaching Assistant\\  \hline
S12	&   26-30/M	& Sighted  	  &     Expert  & Visual Studio Code, Google Colab & Teaching Assistant\\ 
\bottomrule
\end{tabular}
\caption{Participant demographics (programming expertise is self-reported). AT stands for assistive technology.}
\label{table:\partis{}}
\end{table*}

We conducted an IRB-approved study applying \sysname{} design to answer our research questions (Sec.~\ref{sec:rq}). Our study included \nblind{} blind and \nsight{} sighted \partis{} ($23$ males, $1$ female). Through mailing lists, personal outreach, and online programming communities for blind users, we recruited participants who met two key criteria: fluency in English and basic knowledge of \texttt{Python}. 

Our participant pool reflects the persistent gender disparity in computer science~\cite{agarwal2016women}, despite our active efforts to achieve gender balance. Each group included 4 novice and 8 expert Python programmers -- expertise levels that participants self-reported and we verified during the study. Table~\ref{table:\partis{}} presents their demographics. Throughout our analysis, we refer to blind participants as B1-B12 and sighted participants (using \sysname{}) as S1-S12.

\subsection{Apparatus and Stimuli: Coding Tasks and Tools}
Participants completed three types of tasks that represent common programming practices:

\begin{itemize}
    \item[\textbf{T1}] \textit{Code Reading and Understanding}: For a given code snippet, participants predicted its output and described the context of a statement (e.g., its current level, the maximum level of the code, and the context hierarchy).
    \item[\textbf{T2}] \textit{Error Correction}: For a given code snippet containing a syntax error, participants located and corrected it.
    \item[\textbf{T3}] \textit{Code Writing}: Participants implemented a given pseudocode in Python and ran it.
\end{itemize}

\begin{figure}[!t]
  \centering
  \begin{subfigure}[]{0.8\linewidth}
      \centering
     \includegraphics[width=\linewidth]{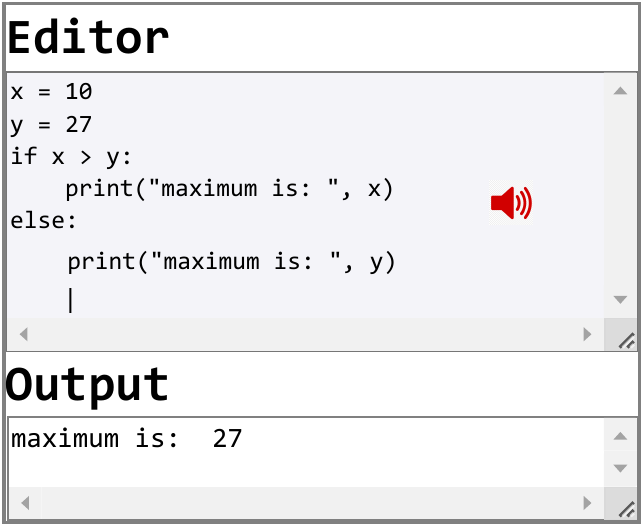}
      \Description{}
  \end{subfigure}
  \caption{A screenshot of our simple, accessible IDE used in the study. It has two main components: the \texttt{\small Editor} (top) and \texttt{\small Output} panel (bottom). It provides TTS output that matches standard screen reader behavior, indicated by the red speaker icon. 
  }
  \Description{A screenshot of our simple, accessible IDE interface divided into two stacked rectangular panels.
The top panel is labeled Editor and contains Python code with syntax highlighting. The code displays a simple conditional statement comparing two variables: x = 10 and y = 27. The if-else statement determines which value is larger and prints it as the maximum. A vertical blue cursor line is visible in the code at the end of the else block. A red speaker icon appears on the right side of the editor, indicating text-to-speech functionality.
The bottom panel is labeled Output and shows the result of executing the code: maximum is: 27. This demonstrates that the else branch was executed since y (27) is greater than x (10).
Both panels have navigation controls, including scroll arrows on the sides and additional utility buttons in the corners. The entire interface has a clean, minimal design with a light background and dark text for readability. The figure illustrates the accessible IDE used in the study, with its text-to-speech capability (shown by the speaker icon) that mimics standard screen reader behavior.}
  \label{fig:system}
\end{figure}

We presented these tasks sequentially from T1 to T3 as their complexity increased. Participants used our browser-based, simple \texttt{\small IDE}, shown in Fig.~\ref{fig:system}, that mirrors the basic features of Notepad++, the most common editor for blind programmers~\cite{Albusays2017Interviews}.
The IDE includes an accessible \texttt{\small Output} panel where participants could run their Python code and view results.
It supports three key operations: executing code and viewing output with \texttt{\small Ctrl + O}, returning to the editor with \texttt{\small Ctrl + E}, and navigating directly to specific lines using \texttt{\small Ctrl + G} through a browser pop-up. It provides text-to-speech (TTS) feedback at all times.

\subsection{Study Procedure}
We conducted sessions in two settings: remotely over Zoom with blind participants and in-person in a quiet office room with sighted participants, following our \sysname{} setup (Fig.~\ref{fig:teaser}.3). After collecting consent and demographic data, blind participants briefly turned on their cameras to verify vision status, and we shared our text editor's URL with them through email or chat.

The study began with a training session introducing our IDE's features, shortcuts, and Python basics. Participants learned essential text navigation shortcuts for TTS interaction, including character and line reading. They practiced basic text editing with TTS feedback: writing lines, browsing content, and removing characters or lines as needed. Once comfortable with the interface, participants wrote simple Python code -- declaring variables, printing values, performing arithmetic operations, and creating if statements -- to familiarize themselves with the programming environment.

After participants mastered the shortcuts and training materials, we began the task sequence. Each task included two to three trials with minor variations to maintain consistent difficulty. Before each trial, we explained its requirements and confirmed participants' understanding.
We instructed participants to maintain their regular pace while prioritizing accuracy.
A trial was completed when participants stated their completion. If participants failed to complete a trial within 5 minutes, we marked it as a failure and removed the data from our performance calculations. We noted three such failures. We addressed participants' questions between tasks.

Two researchers conducted the study.
To observe participant interactions, we asked blind participants to share their screens while coding, and for sighted participants, one researcher monitored their activities on a separate screen not accessible to the participants. We recorded all sessions for later analysis.

Following task completion, we conducted exit interviews to understand observed behaviors and gather feedback about the IDE. Each session lasted 60-90 minutes, and participants received compensation at USD \$25 per hour. 

\subsection{Data Collection, Processing, and Analysis}
Our analysis combined qualitative and quantitative approaches to understand participants' programming strategies. From recorded sessions, we transcribed interviews and analyzed participants' programming behaviors, focusing on their task-solving approaches and navigation patterns through cursor movements. We integrated these observations with experimenters' notes to build a comprehensive view of participant strategies.

We measured three quantitative metrics (see Table~\ref{tab:performance}):
\begin{itemize}
    \item Task completion time (in seconds), averaged across multiple trials for T1 and T3.     
    \item Code understanding accuracy in T1, based on participants' ability to identify correct indentation levels and context hierarchies. Participants earned full points for correctly identifying both indentation levels and complete context hierarchies, with point deductions for each missing or incorrect statement in the hierarchy.
    \item Error count in T3, tracking mistakes made before achieving the desired output.
\end{itemize}

Our qualitative analysis followed an iterative coding process~\cite{bryman1994analyzing}. The first iteration identified low-level descriptive codes that captured specific programming strategies for each task. Later iterations consolidated these codes into higher-level categories, revealing core programming strategies. Through weekly research meetings, all authors refined the codebook's themes and resolved any conflicts.

%% file: 4_findings.tex
\section{Findings}
We now report the task performance and coding strategies of blind and ToPSen \partis{} that emerged from our study.

\begin{table*}[!t]
\small
\begin{tabular}{L{4.3cm} | L{3.8cm} | L{1.3cm} | L{1.3cm} | L{1.3cm} | L{1.3cm}}
\multirow{2}{*}{\textbf{Task}}	& \multirow{2}{*}{\textbf{Metric}} &\multicolumn{2}{c | }{\textbf{Blind}} &  \multicolumn{2}{c}{\textbf{ToPSen}} \\
            	&              & \textbf{Expert} & \textbf{Novice} & \textbf{Expert} & \textbf{Novice} \\
\toprule                  
\multirow{3}{*}{T1: Code Reading and Understanding}	& Completion Time (in seconds) \newline \textit{the lower the better} & $\mu:$ \textbf{140.6} \newline $\sigma:$ 23.84 & $\mu:$ 190.89 \newline $\sigma:$ 82.61 & $\mu:$ 181.38 \newline $\sigma:$ 50.17 & $\mu:$ 225.17 \newline $\sigma:$ 60.77\\ \cline{2-6}
	& Accuracy (\%) \newline \textit{the higher the better} & $\mu:$ \textbf{79.65} \newline $\sigma:$ 10.91 & $\mu:$ 46.25 \newline $\sigma:$ 20.21 & $\mu:$ 77.4 \newline $\sigma:$ 11.15. & $\mu:$ 41.65 \newline $\sigma:$ 5.54 \\ \midrule
T2: Error Correction & Completion Time (sec) & $\mu:$ \textbf{71.67} \newline $\sigma:$ 38.78 & $\mu:$ 106.5 \newline $\sigma:$ 52.02 & $\mu:$ 96.45 \newline $\sigma:$ 54.54 & $\mu:$ 237.33 \newline $\sigma:$ 64.30\\ \midrule
\multirow{3}{*}{T3: Code Writing} & Completion Time (sec) & $\mu:$ \textbf{81.33} \newline $\sigma:$ 45.73 & $\mu:$ 181.5 \newline $\sigma:$ 51.47 & $\mu:$ 131.5 \newline $\sigma:$ 48.91 & $\mu:$ 254.83 \newline $\sigma:$ 84.67\\ \cline{2-6}
		& Number of Errors/Block (count) \newline \textit{the lower the better} &  $\mu:$ \textbf{0.40} \newline $\sigma: $ 0.89 & $\mu:$ 2.75 \newline $\sigma:$ 1.71 & $\mu:$ 1.50 \newline $\sigma:$ 1.70 & $\mu:$ 5.33 \newline $\sigma:$ 1.53 \\\bottomrule
\end{tabular}
\caption{Participants' coding performance metrics (mean and standard deviation) across reading (T1), error correction (T2), and writing (T3) tasks.}
\label{tab:performance}
\end{table*}

\subsection{Task Performance}
Expert blind \partis{} outperformed all other \partis{} across all tasks, demonstrating their ability to manage both intrinsic and extraneous cognitive load. Table~\ref{tab:performance} shows the performance metrics of blind and ToPSen \partis{} grouped by expertise. 
Expert ToPSen \partis{} achieved comparable accuracy to their blind counterparts in comprehension tasks (79.65\% vs. 77.4\%), though they required approximately 30\% more time due to the extraneous cognitive load of auditory perception. All novice \partis{} performed below expectations, achieving less than 50\% accuracy in comprehension tasks while requiring substantially more time to complete them.
The performance pattern showed expert blind \partis{} completing tasks most quickly, followed by expert ToPSen, novice blind, and novice ToPSen \partis{} respectively. However, novice programmers showed high variability in their performance, with considerable spread in their metrics.


\subsection{Developing a Mental Model of Code Snippets through Auditory Channel}
\label{sec:code-overview}
Due to the transient nature of audio, all \partis{} listened to code snippets multiple times to develop a mental model. They controlled the audio readout by moving the cursor -- up/down movements read entire lines, while left/right movements read the character after the cursor position.

Blind \partis{} consistently read code from top to bottom to construct a high-level overview, regardless of task requirements. During this initial pass, they focused on structural elements (e.g., number of lines, declared variables), identified block statements and their types (e.g., \texttt{\small if}, \texttt{\small for}), and grasped the code's overall purpose.

Expert blind \partis{} particularly examined characters to understand code levels and parent-child relationships between statements. B7, an expert blind participant, described this approach:

\begin{quote}
\emph{``Usually, the attempt I take at first is just to read line by line. With this attempt, most things usually become quite clear, except for a few ambiguities, like whether a statement is nested within another statement or not. So, I check one or two lines in more detail \emph{[character by character]}.''}
\end{quote}

Expert blind \partis{} demonstrated the ability to process more than 10 statements and 4-5 levels through a quick overview. Two expert blind participants successfully worked with code snippets of 15 statements and 6 levels of nesting, accurately determining the context of specific statements.

ToPSen \partis{}, in contrast, adopted a more targeted approach. Their cursor movements revealed they focused on understanding each statement in isolation. Except for S7, they tended to memorize constant values (e.g., items in lists or arrays) through character-by-character listening, \fx{pausing at block statements (e.g., \texttt{\small if}, \texttt{\small else}, and \texttt{\small for} statements)}, and cycling their cursor within blocks before proceeding.
Compared to expert blind \partis{}, ToPSen expert \partis{} showed a limited capacity to process code via audio, working with 6-8 statements simultaneously and tracking up to 4 levels.

\fx{
Experts from both groups employed a talk-out-loud strategy, verbalizing logical statements to focus on understanding the code's purpose. We hypothesize that talking out loud might help them, especially experts in the \sysname{} group, to store more information in their two main temporary storage systems in the brain's working memory for auditory information: the phonological loop and echoic memory for 2 to 4 seconds~\cite{baddeleyj}.
}

Novice \partis{} -- both blind and \sysname{} groups -- required multiple (3-5) listens when code exceeded 5-6 statements. They consistently struggled with tracking code levels beyond 3 levels and had difficulty considering parent-child relationships when developing mental models.
Lack of verbalization indicated that they were having difficulty developing mental models. For example, S3 and B5 moved the cursor character by character without verbalizing, often listening to partial statements before moving to different sections. When asked about their behavior, they reported feeling confused.

In summary, expert blind programmers adopt a global approach to code comprehension, while expert \sysname{} programmers rely on local understanding. This contrast likely stems from expert blind programmers' enhanced capacity to process multiple audio chunks in working memory, while \sysname{} programmers, less accustomed to audio as their primary input channel, find maintaining global comprehension through audio cognitively demanding. 
\fx{
Nonetheless, experts in both groups attempted to store more information in their temporary storage systems (working memory) by talking out loud. Novice programmers in both groups struggled to create mental maps of code snippets despite multiple listening attempts.
}

\begin{figure*}[t!]
    \centering
    \begin{subfigure}{\linewidth}
        \centering
        \includegraphics[width=\linewidth]{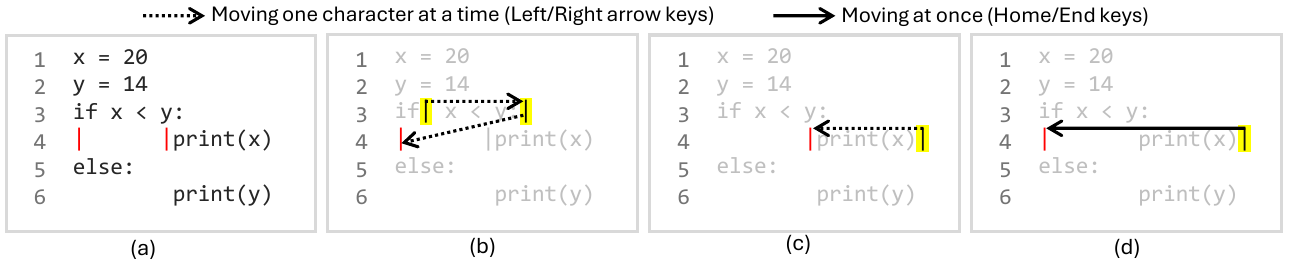}
    \end{subfigure}
	\vspace{-10pt}
 \caption{(a) The two anchor points to count the number of spaces. (b) - (d) Three different ways to reach anchor points.}
	\label{fig:anchors}
    \Description{A diagram showing Python code navigation techniques across four code editor panels labeled (a) through (d). Each panel displays the same Python code snippet with line numbers 1-6 containing variables x=20 and y=14, followed by an if-else statement.
Panel (a) shows the initial state with all text in black and two red vertical cursor marks: one at the beginning of line 4's indentation and another before "print(x)".
Panel (b) illustrates "Moving one character at a time" using Left/Right arrow keys, showing navigation between anchor points. The code outside the navigation path is grayed out, with yellow highlighting on the current cursor positions. Dotted arrows show the step-by-step movement between the characters.
Panel (c) and (d) demonstrate "Moving at once" using Home/End keys, with panel (c) showing an intermediate position and panel (d) showing the complete movement. Both panels use solid arrows to indicate direct jumping between positions, with yellow highlighting marking destination points and red vertical bars marking the starting points.
The top of the figure contains two navigation method labels with corresponding arrow styles: dotted arrows for character-by-character movement and solid arrows for immediate jumps. The figure illustrates how blind programmers use different keyboard navigation techniques to reach anchor points for determining indentation levels in Python code.}
\vspace{-10pt}
\end{figure*}

\subsection{When and Why Programmers Resort to Character-Level Inspection}
Beyond the basic line-by-line navigation, \partis{} switched to character-by-character inspection in specific scenarios: examining list elements, verifying logical operators in block conditions (e.g., the modulo operator \%), and distinguishing variables from fixed strings in \texttt{\small print} statements (e.g., \texttt{\small print(num, `` is prime")}).
\fx{
The most important use of character navigation emerged in determining statement indentation levels in Python, which TTS skipped during line-by-line reading. This process required \partis{} to establish `anchor' points -- which indicates the line's first character (left anchor) and the last space before the statement (right anchor), encapsulating all the whitespace characters constructing indentations. Fig.~\ref{fig:anchors}(a) shows these defined anchor points as two red vertical cursors at line 4.}
\Partis{} used three strategies to reach these anchors: (i) navigating from the previous line's end (Fig.~\ref{fig:anchors}(b)), (ii) moving left from the statement to find the right anchor (Fig~\ref{fig:anchors}(c)), or (iii) using Home/End keys to reach the left anchor directly (Fig.~\ref{fig:anchors}(d)). 

In summary, while prior work noted the tedious nature of counting indentation~\cite{Albusays2016Eliciting, Albusays2017Interviews}, our findings revealed that locating these anchor points added another layer of cognitive complexity to the task. This also suggests that future audio-based programming environments should provide direct mechanisms for conveying code structure without requiring manual character counting.

\subsection{Comprehending Code to Predict the Output}
When predicting output from code containing a \texttt{\small print} statement, \partis{} followed a two-phase approach: first building a mental model of the code's logic, then applying this logic to determine the output. Their cursor movements revealed distinct strategies between expert and novice programmers.

Expert blind \partis{} demonstrated efficient comprehension -- half grasped the logic during their initial overview, moving their cursor directly to the variables for output prediction. The remaining experts examined block statements immediately after the overview before proceeding to variables. 

In contrast, novice blind \partis{} restarted their examination from the first line after reaching the \texttt{\small print} statement. As B4 noted, \textit{``This is the print statement, right? Let me start from the beginning again.''} This behavior suggests that expert blind \partis{} could process more information during a single overview than the novice blind \partis{}.

ToPSen \partis{} approached comprehension through focused examination of individual statements, as mentioned in Sec.~\ref{sec:code-overview}. Some attempted early data memorization -- S5 created memory aids for the list \texttt{\small [-7, 1, 9, -5, 4]} by grouping numbers: \textit{`719 and 54, okay'}. S3, after repeatedly listening to list elements, expressed frustration: \textit{`How can I remember all of these values!'} These initial memorization attempts proved redundant as they needed to revisit the values after understanding the logic. S7's verbalization illustrates their eventual two-phase strategy:

\begin{quote}
\emph{``Oh! I understand what's happening in the code. So, there is a list of numbers at first. We have a sum variable set to zero. Then, we go through each element of the list, and if it is greater than zero, we add it to the sum. Finally, we print the sum variable. So, the code will print the sum of the positive numbers in the list. Let's see... \emph{[moving the cursor back to the list and listening to each list element 2 times]} ...the output will be 15.''}
\end{quote}

While all expert \partis{} accurately predicted outputs, only one novice ToPSen participant (S3) succeeded. Novice \partis{} struggled with nested conditions and tracking variable updates through loop iterations.
\fx{
B5's misinterpretation of Python's \texttt{\small for ... in} loop syntax -- assuming indices rather than list elements (for example, the loop variable \texttt{\small i} gets the values 5, 4 and 3 in three iterations for the statement: \texttt{\small for i in [5, 4, 3]}) -- highlights the conceptual challenges novices face.
}

In summary, programming expertise, not auditory processing experience, determines successful code comprehension through audio. Expert programmers efficiently manage both programming concepts and auditory input, while novices struggle primarily with fundamental programming constructs regardless of their audio processing capabilities.

\begin{figure*}[t!]
    \centering
    \begin{subfigure}{\linewidth}
        \centering
        \includegraphics[width=\textwidth]{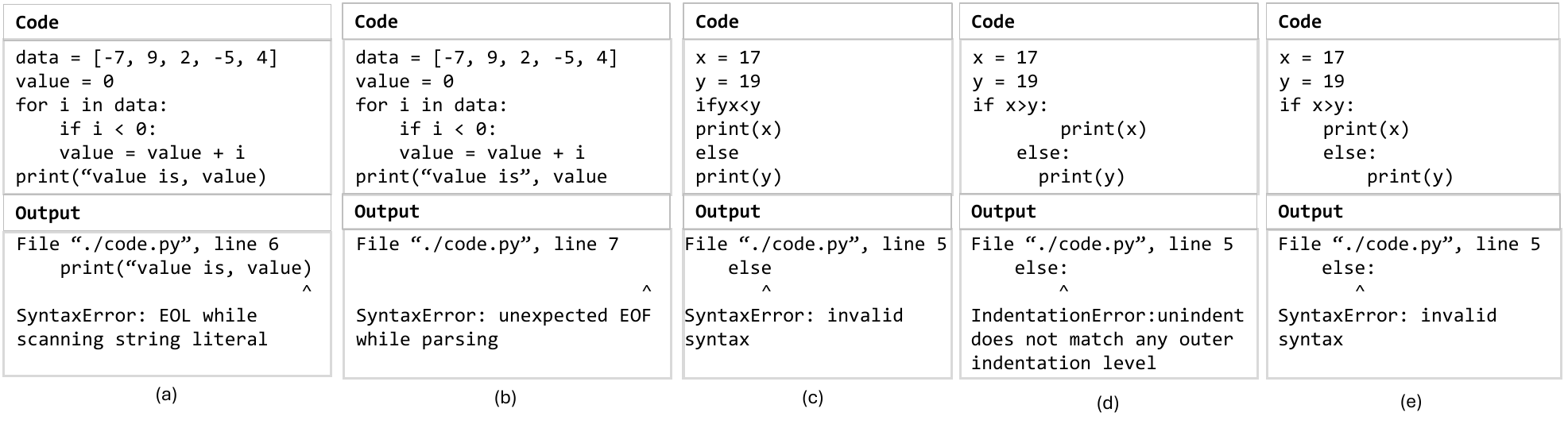}
        \label{fig:error-a}
    \end{subfigure}
    \vspace{-13pt}
	\caption{Five code snippets written by the participants and the associated errors, separated by a dashed line. In all error messages, the visual indicator indicates the location of the error. In the first two snippets, the only difference is the missing `)' at the last line, which changes the error message. In the second snippet, the statement with the error is also missing.
	In the last three snippets, the indentation of the \texttt{\small else} statement and the colon (:) are different, which changes the error message. 
	}
	\label{fig:compiler-errors}
\vspace{-10pt}
\Description{A figure showing five code editor panels labeled (a) through (e), each containing Python code snippets with syntax errors and their corresponding error messages. Each panel is divided horizontally into a "Code" section above and an "Output" section below.
Panels (a) and (b) contain similar code working with a list of integers, but both have syntax errors in the print statement. Panel (a) shows a missing closing parenthesis with the error "SyntaxError: EOL while scanning string literal". Panel (b) shows both a missing parenthesis and the entire statement with the error, resulting in "SyntaxError: unexpected EOF while parsing".
Panels (c), (d), and (e) contain identical code comparing variables x=17 and y=19, but with different indentation patterns and syntax formatting for the if-else statement. Panel (c) shows missing spaces between operators in "ifx<y" with "SyntaxError: invalid syntax". Panel (d) displays proper syntax but incorrect indentation with "IndentationError: unindent does not match any outer indentation level". Panel (e) shows similar code with another "SyntaxError: invalid syntax".
In all cases, the output section displays the filename, line number, and location of the error with a caret (^) symbol indicating the exact position where the error was detected. The figure illustrates how similar code errors produce different error messages, highlighting the challenges novice programmers face in interpreting error messages, particularly when the visual indicators for error location might not be accessible through screen readers.}
\end{figure*}

\subsection{Error Detection Strategies}
We provided a code snippet with a syntax error and allowed \partis{} to execute the code at their discretion. Nine out of twelve blind \partis{} opted to overview the code first, as described in Section~\ref{sec:code-overview}. When offered the option to run the code, B5 requested to review it first. While B1 initially executed the code to identify the error, she examined the code before implementing fixes. She explained her approach:

\begin{quote}
\emph{``I like it this way. I know about the error, but nothing about the code I am going to fix. I need to understand first what the code is doing. Before fixing the error, I want to know what I am getting into.''}
\end{quote}
Two expert blind \partis{} (B3 and B7) detected and fixed the error during their review without executing the code. B1, B3, and B7 also searched for semantic errors.

In contrast, all \sysname{} \partis{}, except for S4, S6, and S12, favored executing the code for error detection. Expert \sysname{} \partis{} preferred using the \texttt{\small Go To} feature to navigate directly to the error line. S7 prioritized speed, stating that reading the entire code was unnecessary for fixing a syntax error. S3 executed the code to avoid character-by-character inspection for subtle errors like missing brackets (as shown in the last line of Fig.~\ref{fig:compiler-errors} (b)).

Despite their preference for code execution, ToPSen \partis{} struggled to recall error structure and navigate error messages. As shown in Fig.~\ref{fig:compiler-errors}, syntax errors display four distinct lines of information: file name with line number, erroneous statement, visual indicator {\large $\hat{}$} pointing to the error, and error type with a brief message. ToPSen \partis{} showed confusion when encountering the visual indicator (announced as ``Caret'' by TTS), indicated by S3's question: \textit{`Shouldn't it say the line number?'} All ToPSen \partis{} needed experimenters' assistance to navigate the error output.

All novice \partis{}, except S6, required code execution and error message review to detect syntax errors. Even B5, who checked all statements character by character, could not identify the issue without executing the code.

In summary, expert blind and ToPSen programmers demonstrate contrasting approaches to error detection and resolution. This distinction highlights how early training and sensory preferences influence debugging strategies: blind programmers develop robust code review habits through their reliance on audio, while ToPSen programmers, accustomed to visual debugging, find navigating audio error messages challenging without visual reference points.

\begin{figure*}[t!]
    \centering
    \begin{subfigure}{\linewidth}
        \centering
        \includegraphics[width=0.8\linewidth]{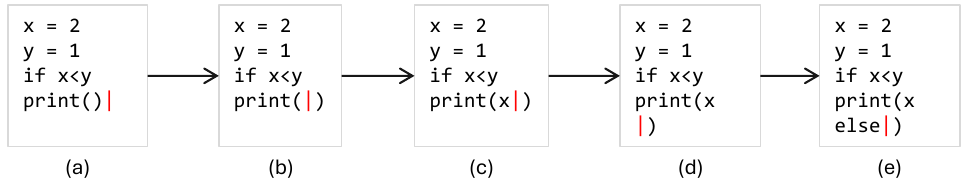}
    \end{subfigure}
	\vspace{-10pt}
 \caption{A sequence of editing steps leading to a deviation from the mental model. The cursor at each editing step is represented using a red vertical line. (a) The participant writes the \texttt{\small print} statement and writes opening and closing brackets first to match the pair. (b) The participant moves left to place the cursor inside the brackets. (c) The participant types the argument of the \texttt{\small print} statement. (d) The participant presses \texttt{\small Enter} to move to the next line without checking the cursor location. The closing bracket moves to the next line without the participant having any clue. The TTS does not provide any cue to indicate this. (e) The participant continues writing the code, and the error continues to move away, deviating from the participant's mental model. As a result, the participant faces difficulty figuring out the issue when they encounter an error.}
	\label{fig:error-moving}
    \Description{A sequence of five code editor panels labeled (a) through (e), showing the progressive development of Python code and how errors can occur during the editing process. Each panel contains the same initial code variables (x = 2, y = 1) and conditional statement (if x<y), with a red vertical line indicating the cursor position at each stage.
Panel (a) shows the beginning of code creation with a print statement containing empty parentheses and the cursor positioned after the closing parenthesis.
Panel (b) shows the cursor moved inside the empty parentheses of the print statement, preparing to add content.
Panel (c) displays the addition of the variable 'x' as an argument within the print function, with the cursor positioned after 'x' but before the closing parenthesis.
Panel (d) illustrates what happens after pressing Enter: the closing parenthesis has unexpectedly moved to the next line, with the cursor now positioned before it. This creates an unintended code structure that doesn't match the programmer's mental model.
Panel (e) shows further deviation from the intended structure, with the programmer having added "else" before the closing parenthesis that was inadvertently moved to a new line. This creates syntactically invalid code that won't execute as intended.
Black arrows between each panel indicate the sequential progression of the editing process. The figure illustrates how blind programmers can develop inconsistencies between their mental model and the actual code structure when text-to-speech output doesn't provide adequate cues about formatting changes that occur during editing.}
\vspace{-10pt}
\end{figure*}

\subsection{Mental Model Verification Strategies}
\label{sec:writing}

Writing code through an auditory interface presents unique challenges in maintaining synchronization between programmers' mental models and the actual code in the editor. Since text-to-speech (TTS) only announces information upon cursor movement, undetected errors can accumulate as programmers continue writing code, creating error cascades. Fig.~\ref{fig:error-moving} illustrates a typical error cascade: B4 wrote a \texttt{\small print} statement with matching parentheses but lost track of the closing parenthesis while editing. This led to multiple errors: missing colon, incorrect indentation, and misplaced parentheses -- all unnoticed as the cursor moved away from the error locations.

Expert blind programmers demonstrated robust verification strategies through constant cursor awareness and systematic review processes to catch errors early, thereby preventing such cascades.
They verified cursor position before pressing \texttt{\small Enter} to prevent statement splitting and checked indentation levels before writing new lines. Their verification process included frequent reviews of 1-2 prior statements for correctness, with some performing complete code reviews before execution. These programmers consistently double-checked cursor location after any position changes, ensuring their mental model matched the actual code.

Expert ToPSen programmers, in contrast, prioritized logical correctness over positional awareness. They generally maintained syntax while focusing on logic but struggled with multi-level indentation changes. While four out of eight reviewed code logic and levels, they typically checked levels after writing logic rather than before. These programmers rarely verified cursor position during out-of-place edits, though one expert (S7) successfully prevented errors through talking out loud and carefully counting indentation. None performed complete code reviews after editing, suggesting less emphasis on maintaining precise cursor awareness and error cascade prevention.

Novice programmers in both groups showed limited verification practices and faced significant challenges maintaining syntax while focusing on logic. Blind novices checked cursor location before editing, but not after, while ToPSen novices wrote code linearly without cursor movement. 
They rarely reviewed statements, frequently missed colons and indentations, and often split statements or wrote multiple statements on one line.
ToPSen novices struggled with both cursor awareness and syntax, indicating the compounded difficulty of managing multiple aspects of code writing through an auditory interface. Only one ToPSen novice (S3) uniquely checked code by listening to statements, demonstrating an awareness of the importance of verification and error prevention.

\subsection{Personal Coding Style}
Blind and ToPSen programmers reveal distinct approaches to maintaining coding style when working through auditory interfaces. These differences illuminate how sensory modalities shape coding practices and habits.

Blind programmers consistently preserved their established coding conventions throughout their work. B5 maintained his preference for condensed variable declarations, combining multiple assignments on single lines (e.g., \texttt{\small x, y = 5, 10}). These programmers deliberately controlled spacing in their statements (e.g., \texttt{\small x = 5} versus \texttt{\small x=5}), demonstrating awareness of code aesthetics even through audio. During code reviews, experts like B7 actively refined stylistic elements, transforming inconsistent spacing such as \texttt{\small result =0} into their preferred format \texttt{\small result = 0}.

Expert ToPSen programmers attempted to maintain consistent spacing but rarely verified these stylistic choices, leading to unintended variations.
Novice ToPSen programmers, focused primarily on achieving functional code through trial and error, paid little attention to stylistic elements.

The shift from visual to auditory coding revealed how deeply visual habits influence the code of ToPSen programmers. S10 reflected that while he typically uses blank lines to separate logical code groups when coding visually, this organizational strategy never occurred to him when working through audio. Other ToPSen programmers acknowledged overlooking spacing inconsistencies around operators (e.g., \texttt{\small count = count +1}) that would have immediately caught their attention in visual coding.

These patterns demonstrate how a programmer's primary sensory mode shapes their coding style. Blind programmers, who routinely process code through audio, seamlessly integrate stylistic considerations into their work. ToPSen programmers, whose coding styles depend on visual feedback, struggle to maintain these practices when working solely through audio. This contrast reveals how coding style becomes embedded in a programmer's primary mode of sensory interaction with code.

%% file: 5_discussion.tex
\section{Discussion}
Our findings reveal that while both expert blind and ToPSen programmers build accurate mental models of syntax and semantics, they differ significantly in how they process structural and positional information. 

Expert blind programmers actively maintain structural information as part of their mental model. They consciously track cursor position, code indentation levels, and output structure, integrating these elements into their programming workflow. These programmers have developed specialized strategies to manage both programming concepts and audio navigation simultaneously, as evidenced by their systematic code review practices and consistent styling habits.

ToPSen programmers, in contrast, demonstrate less awareness of structural elements when working through audio. Accustomed to processing structural information visually, they rarely seek it explicitly unless needed for specific tasks like understanding parent-child relationships. Their difficulty in recalling error message structure and maintaining cursor awareness suggests that information typically processed visually becomes challenging to track when available only through audio.

This contrast likely stems from how each group developed its programming expertise. Blind programmers have evolved strategies that explicitly incorporate structural information into their mental models, treating it as essential rather than supplementary. ToPSen programmers, having developed their skills with visual feedback, treat structural information as background detail -- readily available when needed but not actively maintained in their working memory~\cite{rensink1997, risko2016}. Consequently, when limited to audio, ToPSen programmers focus primarily on code logic while struggling with aspects like cursor position and indentation that their visual system typically processes automatically.

These differences in mental models and information processing suggest potential challenges for mixed-ability programming collaboration. The information blind programmers actively seek and maintain often represents what sighted programmers process passively through visual feedback, creating a potential communication gap and shared understanding during collaborative tasks.

\subsection{Supporting Auditory Programming}
\fx{
Our study indicates that programming expertise significantly influences the mental model development of auditory programmers. Expert programmers create more accurate and robust mental models combining structural, positional, and logical information. Novice programmers, however, face difficulty processing such rich information in their working memory, which leads to inaccurate mental models. Therefore, their difference in expertise demands different needs from tools and IDEs, which we discuss below.
}

\subsubsection{\textbf{Supporting Novice Programmers}}
\fx{
Our research revealed that novice programmers struggle with both the intrinsic and extraneous cognitive load of programming, hindering their ability to construct coherent mental models of code. During reading activities, they face challenges interpreting syntax, identifying hierarchical relationships, and following logical flow. When editing code, they frequently introduce syntax errors, leave errors when moving to new lines, and have difficulty comprehending compiler error messages. Consequently, providing targeted assistance for various coding tasks and facilitating the development of consistent mental models is crucial for them.
}

\fx{
Augmented and structured programming representations can significantly enhance novice programmers' ability to overcome coding challenges. Tree-based code representations~\cite{baker2015struct, Schanzer2019Accessible} help clarify structural and logical relationships, while specialized tools like CodeTalk~\cite{Potluri2018CodeTalk} and Sodbeans~\cite{stefik2009sodbeans} provide auditory support for debugging. For editing tasks, IDEs that constrain unstructured editing prevent syntax errors and foster consistent mental models. Structured environments such as accessible blocky~\cite{Mountapmbeme2022accessibleblockly} enable valid code creation through compatible puzzle pieces representing code blocks. Similarly, grid editor~\cite{ehtesham2022gridcoding} organizes code using rows and columns to represent statements and levels, encoding contextual information within statement rows. These structured approaches help novices maintain cursor positioning and restrict writing to valid code levels.
}

\fx{
Furthermore, IDEs should proactively provide auditory feedback to novice programmers about their cursor location and syntax errors during code editing. Beginners often edit without checking cursor position, inadvertently creating errors when manipulating statements—for example, pressing \texttt{\small Enter} to split statements. Real-time auditory cues about cursor location and potential errors can prevent inconsistencies in their mental models and boost confidence, particularly for programmers who rely primarily on auditory information.
}

\fx{
Beyond editing support, comprehending existing code logic presents a significant challenge for novice programmers. Our findings show that programmers using auditory channels must repeatedly move their cursor over logical statements to understand program logic—a cognitively demanding process where novices often miss essential conditions. Prior research indicates that well-documented code with docstrings and comments can alleviate comprehension difficulties~\cite{pandey2021understanding}. We propose that providing concise logic summaries can further assist programmers in code comprehension, which our \partis{} attempted to facilitate during our study. Recent Large Language Models (LLMs) offer promising capabilities for automatically generating code summaries tailored to novice programmers, though the effectiveness of these tools for supporting beginners requires further investigation.
}

\subsubsection{\textbf{Supporting Expert Programmers}}
\fx{
Expert blind programmers effectively develop and retain mental models of code. They consistently track cursor position, adhere to coding standards, and efficiently interpret and write error-free code. For these advanced users, support extends beyond basic reading and editing assistance to include features that preserve personal coding styles and enhance productivity through modern IDE integrations. Supplementary representations such as tree view~\cite{Potluri2018CodeTalk, baker2015struct} or grid view~\cite{ehtesham2022gridcoding}, coupled with multi-modal interfaces incorporating refreshable braille displays, can significantly enhance their navigation and comprehension capabilities.
}

\fx{
While expert programmers can develop robust mental models, the process requires code overviews through their auditory channel. Our study indicated that expert blind programmers can effectively process approximately 15 statements across 3 to 5 nesting levels, beyond which comprehension becomes challenging. To facilitate understanding, we recommend maintaining code segments within these ranges. For more extensive or complex implementations, decomposing code into multiple methods (functions) proves beneficial—a finding supported by previous research~\cite{pandey2021understanding}, which indicates that shorter code segments simplify navigation for programmers using screen readers and other assistive technologies. This approach not only enhances accessibility but also promotes fundamental software engineering principles such as modularity and single responsibility~\cite{adnan2018software, soft_princ}. Furthermore, method-based code organization enables blind programmers to gain a comprehensive understanding by examining method signatures before diving into implementation details~\cite{armaly2017comparison}.
}

\fx{
Due to their robust mental models, expert programmers may benefit from advanced productivity tools such as auto-completion features. Recent code generation models offer significant potential to enhance their workflow by automating code creation and refactoring tasks, which expert programmers can rapidly validate. Their ability to identify and interpret error messages efficiently allows them to integrate generated code into an existing codebase, substantially improving productivity. The impact of AI-powered
coding tools on blind developers’ effectiveness represents an emerging research domain~\cite{perera2024enhancing, flores2025impact}, which can provide further insights
regarding the effectiveness of such tools to support blind program-
mers’ mental models.}

\fx{
During our study, expert blind programmers also exhibited awareness of cursor positioning, enabling them to prevent errors. However, navigating larger codebases presents challenges when they must seek information in distant locations and subsequently return to their original position. Prior research has identified cursor position loss during backtracking as a significant obstacle for blind programmers~\cite{blinddeveloper, Albusays2016Eliciting}. Consequently, IDE developers should prioritize features that help maintain cursor context and facilitate navigation and backtracking to the editing location, thereby enhancing the productivity of expert blind programmers.
}

\begin{figure*}[t!]
    \centering
    \begin{subfigure}{.8\linewidth}
        \centering
        \includegraphics[width=\linewidth]{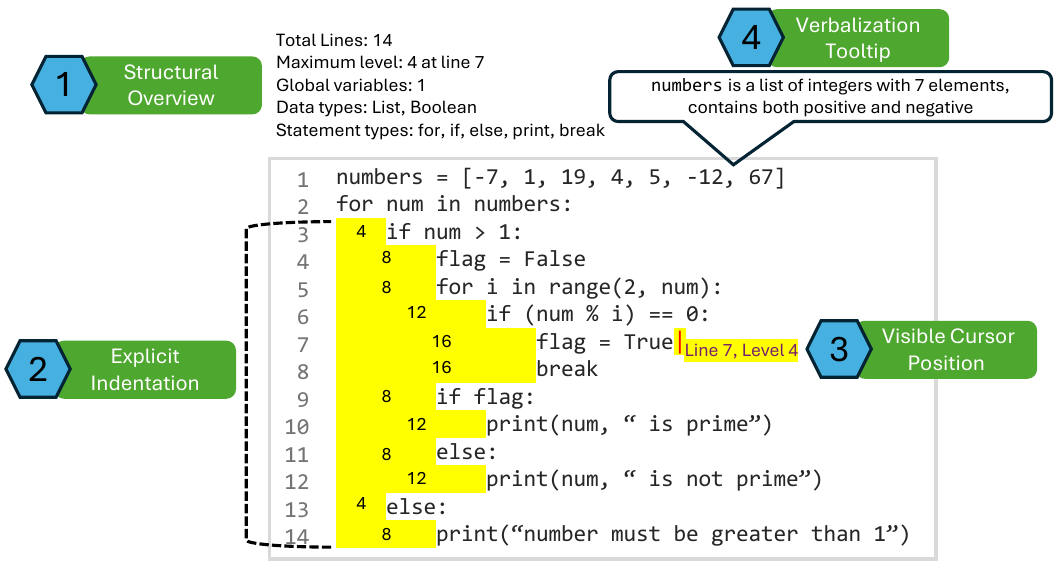}
    \end{subfigure}
	\vspace{-10pt}
 \caption{Potential design to align sighted programmers' mental model with blind programmers during mixed-ability collaboration.}
	\label{fig:sighted-ideation}
    \Description{
    A diagram illustrating a proposed IDE design to bridge the mental models between sighted and blind programmers during collaborative coding. The figure contains a Python code editor with enhanced accessibility features highlighted by four numbered hexagonal elements.
The code sample shows a prime number checking algorithm with 14 numbered lines. Yellow highlighting indicates code indentation levels, with numerical indicators (4, 8, 12, 16) showing the exact number of spaces for each indentation level.
Element 1 "Structural Overview" provides metadata about the code: total line count (14), maximum indentation level (4 at line 7), global variable count (1), data types used (List, Boolean), and statement types (for, if, else, print, break).
Element 2 "Explicit Indentation" points to the left margin of the code where indentation levels are numerically displayed, connected by a dashed line spanning the entire code block.
Element 3 "Visible Cursor Position" shows the current cursor location highlighted at line 7 with a yellow indicator and explicit text "Line 7, Level 4" for precise positioning information.
Element 4 "Verbalization Tooltip" displays contextual information about the variable "numbers" currently in focus, stating "numbers is a list of integers with 7 elements, contains both positive and negative" to aid in code comprehension.
The code itself is a Python script that iterates through a list of integers, checking whether each number greater than 1 is prime or not using a flag-based algorithm with nested loops. The design demonstrates how explicit structural information, cursor positioning, and contextual tooltips can create a shared understanding between blind and sighted programmers in a collaborative environment.
    }
\vspace{-10pt}
\end{figure*}

\subsection{Fostering Mixed-Ability Collaboration}

We identified differences in sighted and blind programmers' mental models, which indicate that, by default, these two groups prioritize different information during coding.
If the gap in the mental model persists, effective communication between sighted and blind programmers will be hampered. Therefore, IDEs should focus on bridging the gap during collaborative sessions and make sure sighted programmers pay attention to the information that blind programmers also need. Fig.~\ref{fig:sighted-ideation} illustrates four potential ideas to bridge the gap in mental models so that sighted programmers can effectively communicate with blind peers.

\fx{
First, IDEs can \textbf{\textit{provide structural summary}} information at the top of the source code that sighted programmers, otherwise, may overlook but are essential for blind programmers, as shown in Fig.~\ref{fig:sighted-ideation}(1).
}
Sighted programmers can quickly communicate this information before jumping into detail during mixed-ability collaboration. Such a practice can draw structural attention to sighted programmers while also benefiting blind programmers to get a quick overview of the code structure without doing it manually.

\fx{
Second, \textbf{\textit{representing visual information explicitly}}, such as indentation with colors and texts, can attract sighted programmers' attention and allow them to focus on code levels that they consume subconsciously.
}
This is similar to screen reader plugins that announce indentation explicitly as `4 spaces'~\cite{nvda_addon_speak_indent}. The idea is demonstrated in Fig.~\ref{fig:sighted-ideation}(2), where both color and text is used to explicitly represent indentations. Similarly, common errors can be shown in different colors to attract sighted programmers' attention.

\fx{
Third, IDEs can adopt \textbf{\textit{visible cursor position}} along with the line and level explicitly tagged with the cursor, as shown in Fig.~\ref{fig:sighted-ideation}(3).
}
As such, sighted programmers will always notice the location and may convey it to blind programmers, who require such detail to develop their mental model.

Finally, prior work indicated that sighted programmers do not verbalize on-screen contents and move around quickly without effectively communicating information~\cite{cha2024understanding}. In addition, they often refer to information that requires visual perception to follow (e.g., this line, that object). Our findings suggest that sighted programmers may not be aware of this fact due to subconsciously consuming information. 
\fx{
IDEs can support in this case by providing \textbf{\textit{verbalization tooltips}} that can guide sighted users to verbalize information effectively, demonstrated in Fig.~\ref{fig:sighted-ideation}(4).
}

We propose that IDEs provide a \textit{collaboration mode} encompassing the above ideas to foster effective collaboration between sighted and blind programmers. Collaborative sessions can also be restricted to keyboard-only navigation for sighted programmers, and blind programmers should be able to follow the cursor of sighted programmers, as suggested in~\cite{potluri2022codewalk}.

\subsection{Applying \sysname{} Paradigm to Other Sensory Modalities}
Researchers can adapt the \sysname{} paradigm to compare performance between groups with different sensory abilities by following a three-step approach.
\textit{First}, design a task where both groups use an identical sensory channel. For example, when comparing hearing and hard-of-hearing participants, researchers could ask both groups to understand dialogue in a movie through the visual channel—reading subtitles, observing lip movements, facial expressions, and body language.

\textit{Second}, apply environmental constraints to the secondary channel where the two groups differ. For hearing and hard-of-hearing participants, this secondary channel is auditory. Rather than simulating deafness with earplugs, researchers could add realistic background noise to the movie audio, such as aircraft engine, drone, or construction sounds, that renders the soundtrack unintelligible for hearing participants. This forces hearing participants to rely on the same visual information that hard-of-hearing participants would naturally use.

\textit{Third}, conduct the study under these controlled conditions. Both groups now process information through the identical visual channel, revealing how each group adapts their comprehension strategies when their differing auditory abilities become irrelevant due to environmental constraints.

By aligning both groups to an identical channel through realistic environmental challenges, the \sysname{} paradigm enables authentic comparisons while avoiding the ethical concerns of traditional disability simulation.

\subsection{Future Work}
Our findings reveal fundamental differences in how blind and sighted programmers process code through auditory channels. These insights suggest two promising research directions that could improve IDE accessibility and support mixed-ability collaboration, such as pair programming between sighted and blind coders.

\paragraph{\textbf{Developing a Standardized Editor Interface}}
We plan to address the tool fragmentation that forces blind programmers into specific development environments. Despite similar functionality, existing IDEs create vastly different user experiences that require users to master unique interaction patterns for each tool. When blind programmers switch between IDEs, they must relearn shortcuts, navigation patterns, and feedback mechanisms, creating an unnecessary cognitive burden.
Our future work will explore standardized APIs—similar to accessibility frameworks like UI Automation in Windows—that enable consistent shortcuts and uniform feedback across editors~\cite{Billah_sinter}. Where vendors fail to provide such interfaces, we will collaborate with the accessibility community to develop plugins that bridge these gaps~\cite{momotaz2023understanding, momotaz2021understanding}.
This standardization would enable uniform interaction experiences, a cornerstone of usable accessibility for blind users~\cite{Billah_ubiquitous, touhidul2023probabilistic}.

\paragraph{\textbf{Integration of Vision-Language Models in IDEs}}
Our research revealed that expert blind programmers excel at structural awareness in their mental models, while sighted programmers struggle with positional and hierarchical information when limited to audio feedback. We plan to explore how Large Multimodal Models (LMMs)—particularly vision-language models with speech input and audio output—could bridge this gap by serving as virtual sighted coding partners for blind programmers.
Our future work will investigate LMMs that provide audio output in natural speech, mimicking human-like personas. For example, these models could incorporate human vocal patterns—vocal fry for syntax errors, sighs for compilation failures—that blur the distinction between human-generated and machine-generated audio, creating more pleasant user experiences~\cite{han2024uncovering}. Since blind users have become early adopters of LMMs for visual scene description~\cite{xie2025beyond, xie2024emerging}, these systems could naturally extend to describe code structure that sighted programmers process visually.
This integration would bridge the communication gaps we identified in mixed-ability collaboration and provide real-time visual structural information that helps align mental models between programmers with different sensory abilities.

\subsection{Limitations}
Our study has several limitations. First, since we conducted our research using Python, some findings may be specific to this programming language.
For example, findings influenced by Python's whitespace-based indentation, such as finding anchor points, counting spaces, and determining the correct indentation levels, are specific to Python.
Such findings may differ for other C/Lisp-style languages that use matching bracket pairs to indicate code levels, where determining the levels may introduce different strategies and require separate IDE support to foster mixed-ability collaboration.
Investigating other programming languages in the future may lead to more generalizable IDE design guidelines.
However, given Python's widespread adoption, we believe our insights remain valuable to researchers in HCI, programming languages, and accessibility.
Additionally, findings influenced by the difference in mental models between sighted and blind programmers are likely to generalize to other languages.
We believe our proposed IDE design to explicitly convey structural overview, cursor position, and verbalization tips will establish effective communication during mixed-ability collaboration regardless of programming language.

Second, despite our conscious efforts, we did not achieve gender balance among participants. This reflects a broader challenge in computer science education, particularly within the blind community, where high access barriers limit the number of individuals who pursue programming.

Finally, while our sample size of 24 participants meets acceptable standards for HCI research involving disability communities, a larger sample would have provided greater statistical power for our quantitative analysis. Our study was also conducted in a single session with limited tasks and in a basic text editor, which may have impacted ToPSen \partis{}' performance. In the future, a longitudinal study with ToPSen programmers can provide more insights regarding how they adapt their behavior and strategies over time. Furthermore, modern IDEs integrated with our proposed design ideas, as well as emerging AI-powered tools (e.g., an LLM agent assisting in communication), can also be studied in real-world mixed-ability pair programming tasks to validate, refine, and improve these IDEs for effective collaboration.

%% file: 6_conclusion.tex
\section{Conclusion}
Our study compared how sighted and blind programmers work under controlled conditions, introducing ToPSen -- a novel approach that reframes auditory programming as a technical requirement for sighted users. Our findings reveal that expert blind programmers can surpass their sighted counterparts when coding with audio feedback, primarily because the two groups engage with code differently. Blind programmers demonstrated heightened awareness during code modification and showed greater precision, resulting in fewer errors. In contrast, sighted programmers under ToPSen developed adaptive strategies, such as verbalizing code levels, to maintain their mental models. However, due to their visual perception, they subconsciously consume structural and positional information, which are essential active components of blind programmers' mental models. 
These insights extend beyond mere performance comparisons. They illuminate how programmers adapt to non-visual coding environments and suggest ways to improve programming tools for all users. Most significantly, our findings offer practical guidance for mixed-ability collaboration and can enable sighted instructors to teach computer programming to blind or newly blind students.